# Searching for Jupiter and Saturn analog planets with potential icy exomoons - a Panspermia perspective

Balázs Bradák*

*The Expansion Initiative - Laboratory of Exo-Oceans, Faculty of Oceanology, Kobe University, 5-1-1 Fukaeminami-machi, Higashinada-ku, Kobe 658-0022, Japan

ABSTRACT

Despite the exponentially growing research interest in the icy satellites of gas giants in the Solar system, sparked by the possibility of evolving life in their subsurface oceans, there are still only a few studies using them as potential analogies in exomoon studies. Considering the possibility of complex organic molecules and microbial life appearing under the ice shell of those satellites in the Solar system, this study investigates the possible analog sources (targeting the potential ice satellite hosting, Jupiter and Saturn-like planets in exoplanet databases) and the transport of such bioaerosols in an attempt to support or contradict Panspermia, a fringe theory about the fertilization of Earth. Along many general parameters of the candidate planets, the host star, and the star system, additional factors thought to be related to Panspermia were also considered (e.g., the evolution of icy satellites, the frequency of impact related ejection, the traveling time from a source, and so on), revealing the following results. Eleven exosystems, with candidate gas giants hosting icy satellites, were found in a database listing more than 5000 exoplanets. The exomoons of the oldest systems (c.a. > 8 Ga; HD 191939, HD 4203, and HD 34445) could have developed rapidly considering the short formation time (~100 Myrs), which may result in the overlap of the putative early biological evolution and asteroid bombardment phase, providing a higher chance to bioaerosol ejection to space due to frequent collisions. However, the direct transfer might have occurred too early, even before our Solar system was formed, which prevented the fertilization of the latter. A longer formation of the exomoons (~3 Gyrs) significantly reduces the chance of ejection into space by asteroid impacts, which become less frequent over time, but increased the chance of arrival in time to the Solar system. Younger systems, such as HD 217107, HD 219828, HD 140901, and HD 156279 (c.a. 4 to 6 Ga), are better candidates of putative microbe source if the short icy satellite formation, along the higher impact and ejection frequency are expected considering direct transport from those systems. In such cases the putative bioaerosol transfer might arrive in time and capable of fertilizing various planetary bodies in the Solar system.

Keywords: icy satellite; exomoon; gas giant; exoplanet; Panspermia

## 1. INTRODUCTION

Panspermia, a fringe theory, considers the existence of various planetary bodies, where life evolved, and where from, later swarmed out (e.g., by hitchhiking asteroids) to the Universe and fertilized star systems along [1]. Regarding the type of the earliest known planets, i.e., gas giants, the possible incubators of life might be their icy satellites (exomoon), using Jupiter and Saturn, and their ocean and possibly microbial life harboring icy satellites (Europa and Enceladus) as an analogy. Proposing icy satellites as incubators of life attempts to fill some part of a research hiatus, considering the small number of icy exomoon studies [2,3], by connecting the study of icy satellites and gas giants of the Solar and exoplanetary systems with potentially similar configurations. Along with Panspermia, some additional theories and observations are introduced below, which suggest that the icy satellites of exo-gas giants might have been functioning as a source of life since the earlier period of the Universe.

i) Water in the early universe. Water might already be abundant in the "first metal enrichment" epoch when Population (Pop) III stars enriched the cosmos with elements during their short but very productive life. It resulted in abundant water in the interstellar medium when the Universe was only approximately 0.3-0.4 Ga old [4]. Regarding the galactic environment, Loeb [5] suggests that the habitable cosmological epoch, in which life could evolve on planetary bodies, dates back much earlier, when the Universe was only around 10–17 Ma old.

ii) The type of the up-to-date identified oldest planets. One of the oldest planets, Psr B1620-26 B (12.7 Ga) and WASP-183 b (14.9 ± 1.7 Ga), are both Jovian planets. Both WASP-183 b and Psr B1620-26 B (half of the size and more than twice as big as Jupiter, respectively) have been orbiting around G-type stars during their history. However, the progenitor star of the latter later became a member of a binary system following the first 10 Gyr evolution [6-8]. Unfortunately, the similarities end here if we look for more analogies between Jupiter and WASP-183 b. The distance of the gas planet from its host star is only 0.05 AU (Mercury: 0.4 AU), which makes the existence of icy satellites harboring the building blocks of life or even various lifeforms most likely impossible and makes Psr B1620-26 B (and further early planets with similar parameters as the gas giants) to the sole candidate with putative icy exomoons, formed out of its circumplanetary disk and may be observed in the future [9-12].

iii) Existence and maintenance of a subsurface ocean on icy exomoons and the possibility of biological evolution. Building blocks of complex organic compounds, observed, e.g., in the geysers (plumes) of Enceladus [13, 14], might have existed at the time of the formation of the earliest satellites [4,5,15,16]. Still, additional factors are needed to keep the subsurface ocean liquid and oxygenated, i.e., provide a habitable environment [17]. External forces, triggered by the host planet and orbital forcing components, such as diurnal tidal flexing [18-20], forced obliquity [20-23], and eccentricity [22,24], along with hybrid and endogenic processes, such as the asynchronous rotation of the ice shell to the tidal torques [25], changes in the ice shell thickness and viscosity [26], currents in the subsurface ocean [27], and hot-spot analog activity [28,29] may/might maintain liquid ocean, and as a result of the previous ones, cryotectonic processes which supports downward material transport [17].

Despite the exponentially growing research interest and popularity of the icy satellites of gas and ice giants in the Solar system, sparked by the possibility of evolving life in their subsurface oceans, there are still only a few studies using them as potential analogies in exomoon studies [3]. This research aims to find star systems with Jupiter and Saturn analog gas giants in existing exoplanet databases, considering their icy satellites as potential life-harboring planetary bodies, and it attempts to evaluate the possibility of direct bioaerosol transport and fertilization of the early Solar system.

## 2. METHODS

### 2.1 Filtering parameters

This study used NASA`s Exoplanet Archive (https://exoplanetarchive.ipac.caltech.edu/), which at the time of the research contained close to 5,300 exoplanets.

Various groups of filtering parameters were applied during the search. Parameters such as planetary mass (M) and radius (R) (two basic parameters commonly used in exoplanet classifications), along with the orbit semi-major axis (a), define the host planet of the putative exomoons. Another group of parameters, consisting of spectral type and stellar age, characterizes the host star in the star system. In addition, star system parameters, such as the number of host stars and planets, were also used.

As a primary selection of candidates, giant planets were filtered based on the literature's definitive R and M values, $R \geq 10R_\oplus$ and $M \geq 95M_\oplus$ [30, 31]. The list of planets needs to be narrowed down further by applying various parameters, summarized in Table 1.

The allocation of planets beyond the ice (frost or snow) line ($R_{ice}$) is significant in keeping the planetary and environmental condition of the icy satellites quasi-permanent. To define the allocation of $R_{ice}$, the following equation was used [32]:

$$T = \left(\frac{L_\odot}{8\pi {R_{ice}}^2 \sigma_{SB}}\right)^{1/4} \quad \text{(Eq.1)}$$

Where T is 150 K, indicating the freezing point of water in the galactic environment, $L_{☉}$ is the luminosity of the central star, $\sigma_{SB}$ is the Stefan-Boltzmann constant, and $R_{ice}$ is the distance of the ice line from the central star.

Despite the growing number of potential ice satellites with subsurface oceans orbiting around the Solar system's ice giants, only the planets similar to Jupiter and Saturn were considered during the search for possible analog star systems. The icy satellites of the two gas giants, mainly but not limited to Enceladus and Europa, the moons which, based on, e.g. the direct observations of their plums, most likely hide subsurface oceans under their ice shells which may contain organic molecules and hopefully various forms of life [13, 33-36].

To limit the maximum distance from the central star of the candidate host planet, the ratio between the orbital semi-major axis of Saturn and the distance of the ice line was used as a limiting parameter (referred to as "Saturn-line"), with the following condition, which needs to be fulfilled:

$$a_{planet}/R_{ice} \leq a_{♄}/R_{iceSol} \qquad \text{(Eq.2)}$$

where $a_{planet}$ is the orbital semi-major axis of the candidate planet, $R_{ice}$ is the distance of the ice line from the central star, $a_{♄}$ is the orbital semi-major axis of Saturn, and $R_{iceSol}$ is the distance of the ice line from the Sun in the Solar system. Such limitations help control the galactic environment variations (and the analogy) around the candidate exoplanets and the gas giants in the Solar system.

The additional parameters, such as the spectral type of the central star (Class G stars, similar to the Sun), the number of central stars (numS = 1), and the number of planets (numP > 1), help to filter the data further and strengthen the analogy between the Solar system and the candidate star systems and the potential planets which may function as host planets of icy satellites, exomoons, with oceans under their ice shell and possibly harboring life.

The last filtering component, the stellar age, indicates the possible role of the icy satellite orbiting around the candidate host planet as the ground zero location of life. Older systems may function as early incubators of life. Referring to the Panspermia theory, those complex organic molecules and lifeforms may be transferred, swarm out of the early Universe, and fertilize other planetary systems.

| Param. | Jupiter | Saturn | Comments |
|---|---|---|---|
| radius($R_\oplus$) | 11.2 | 9.1 | ≥10* |
| mass($M_\oplus$) | 317.8 | 95.2 | ≥95** |
| Ice line - $R_{ice}$ (AU) | 5.2 | 9.5 | Beyond the calculated ice line*** |
| $a_{max}$ (AU) | $a_{planet}/a_{ice} \leq a_♄/a_{iceSol}$ | | Maximum planetary distance from the ice line needs to be similar to Saturn's distance from the ice line in the Solar system. |
| S.type | G2V | | Class G stars |
| numS | 1 | | Sole star systems |
| numP | 8 | | Planetary config. |
| S.age (Ga) | 4.6 | | Older to younger |

**Table 1**. Parameters applied during the search for potential host planets of icy exomoons. *[30], **[31], ***[32] (Eq. 3); abbreviations: $a_{max}$ – maximum orbital semi-major axis of the candidate host planet; numS – number of host stars; numP – number of planets; S.type – spectral type; and S.age – stellar age

### 2.2 About the classification of exoplanets in the filtered database

One of the commonly applied methods during the classification of exoplanets is the mass (M) vs. planetary radius (R) plots (MR plots). These two observable parameters do not simply represent what their name indicates but may carry information about, e.g., the composition and inner structure of the planet and the characteristics of the atmosphere. The construction of theoretical planets with various characteristics, included in their mass and atmosphere parameters (using the equation of state – EOS for various planetary components such as iron, water, olivine, hydrogen, and so on) led to the determination of composition curves (e.g., 50 w% H2O curve), which are plotted in MR diagrams along with observed exoplanets. Based on their mass and radius data, the observed exoplanets are located in various plot regions along various theoretical curves, indicating the planet's composition. This way, MR plots provide a commonly used and acceptable fund for the classification of the planets [30, 37-41]. When determining the exoplanet types, the classification of Zeng et al. [30] and Helled [31] were used in this study.

### 2.3 Additional parameters

This study also uses a plot indicating the relationship between the host star's mass and the accumulated mass of known planets in the Solar and exosystems. The mass of the protosun and the mass loss of the Sun and solar-mass stars were calculated based on various studies [42-45].

In the case of the Solar system, the accumulated planetary mass was calculated in two ways, one containing only the mass of planets and the other all the known mass of planetary bodies. During the calculation of the latter, the following sources were used. The mass of the satellites were find in the Planetary Satellite Physical Parameters database (Jet Propulsion Laboratory, California Institute of Technology, NASA; https://ssd.jpl.nasa.gov/sats/phys_par/), and the Planetary Fact Sheets database (NASA, https://nssdc.gsfc.nasa.gov/planetary/planetfact.html). Calculations about the Oort cloud`s mass can be found in Weissmann [46] and Marochnik et al. [47], about the Kuiper Belt in Gladman et al. [48] and Bernstein et al. [49], and about the Main Asteroid Belt in Pitjeva and Pitjev [50].

## 3. RESULTS AND DISCUSSION

### 3.1 Basic analysis and findings

#### 3.1.1 Filtering

Applying the criteria summarized in Section 2 and Table 1. resulted in 13 findings out of 5272 exoplanets in the used database (Table 2). Four of the 13 exoplanet candidates had to be analyzed carefully (2 exoplanets, marked by red in Table 2) or excluded from the analysis considering the multiple stellar age and mass data related to the same star system appearing in the database. Please note that the HD 4203 system and HD 50499 were excluded from some analyses due to the differences in the stellar mass data. Still, they may be included in some further sections, with a warning about the potential biases. Despite their potential, the other two candidates were excluded from the analysis due to the significant stellar age differences appearing at various exoplanet members of the same star system (HD 75784c and b, and HD 183263c and b). Thus, such errors left 9 or 11 candidates behind. Among the candidates, two star systems, HD 191939 and HD 34445, consist of six planets. Other star systems have fewer, three (one candidate system: HD 204313) or two observed planetary members. Regarding the age of the star systems, four are older than the solar system, with the oldest being 8.7 Ga. The suggested oldest star system and a similar 8.5 Ga old are the ones hosting six planets. Two younger ones are the 5.84 Ga old HD 217107 and the 5.69 Ga old HD 219828 star systems. The age of the star systems, younger than the Solar system, fell between 4.16 and 2.55 Ga (Table 2). Regarding the star systems with questionable stellar age and stellar mass, HD 4203, a two-known-planet system with one of the oldest ages, 8.52 Ga, was recognized.

| Hostname (S.type; numS) | No of pl. | S.age (Ga) | Planet`s name | Planet`s type | HZouter | Ice line |
|---|---|---|---|---|---|---|
| HD 191939 (G9 V; 6) | 6 | 8.7 | HD 191939 b | transitional/sub-Neptune | IN | INSIDE |
| | | | HD 191939 c | transitional/sub-Neptune | IN | INSIDE |
| | | | HD 191939 d | transitional/sub-Neptune | IN | INSIDE |
| | | | HD 191939 e | gas giant | IN | INSIDE |
| | | | **HD 191939 f** | **gas giant** | **OUT** | **BEYOND** |
| | | | HD 191939 g | transitional/sub-Neptune | IN | INSIDE |
| *HD 4203 (G5; 2)* | *2* | *8.52* | *(HD 4203 b)* | *gas giant* | *IN* | *INSIDE* |
| | | | ***(HD 4203 c)*** | ***gas giant*** | ***OUT*** | ***BEYOND*** |
| HD 34445 (G0 V; 6) | 6 | 8.5 | *(HD 34445 b)* | *gas giant* | *IN* | *INSIDE* |
| | | | HD 34445 c | transitional | IN | INSIDE |
| | | | HD 34445 d | sub-Neptune | IN | INSIDE |
| | | | HD 34445 e | sub-Neptune | IN | INSIDE |
| | | | HD 34445 f | transitional | IN | INSIDE |
| | | | **HD 34445 g** | **gas giant** | **OUT** | **BEYOND** |
| HD 217107 (G8 IV; 2) | 2 | 5.84 | HD 217107 b | gas giant / Hot Jupiter | IN | INSIDE |
| | | | **HD 217107 c** | **gas giant** | **OUT** | **BEYOND** |
| HD 219828 (G0 IV; 2) | 2 | 5.69 | HD 219828 b | transitional | IN | INSIDE |
| | | | **HD 219828 c** | **gas giant** | **OUT** | **BEYOND** |
| | | | | | | |
| HD 140901 (G6 IV; 2) | 2 | 4.16 | HD 140901 b | transitional/sub-Neptune | IN | INSIDE |
| | | | **HD 140901 c** | **gas giant** | **OUT** | **BEYOND** |
| HD 156279 (G6; 2) | 2 | 4.1 | HD 156279 b | gas giant | IN | INSIDE |
| | | | **HD 156279 c** | **gas giant** | **OUT** | **BEYOND** |
| HD 66428 (G8 IV (+G); 2) | 2 | 3.515 | HD 66428 b | gas giant | OUT | INSIDE |
| | | | **HD 66428 c** | **gas giant** | **OUT** | **BEYOND** |
| HD 204313 (G5 V; 3) | 3 | 3.38 | HD 204313 b | gas giant | OUT | INSIDE |
| | | | HD 204313 c | transitional/sub-Neptune | IN | INSIDE |
| | | | **HD 204313 e** | **gas giant** | **OUT** | **BEYOND** |
| HD 92788 (G6 V; 2) | 2 | 2.55 | HD 92788 b | gas giant | IN | INSIDE |
| | | | **HD 92788 c** | **gas giant** | **OUT** | **BEYOND** |
| *HD 50499 (G1 V; 2)* | *2* | *2.4* | ***HD 50499 c*** | ***gas giant*** | ***OUT*** | ***BEYOND*** |
| | | *2.391* | *HD 50499 b* | *gas giant* | *OUT* | *INSIDE* |

**Table 2**.  Jupiter and Saturn analog exosystems and their known planets, including the candidate gas giants (bold letters) found in the database following the filtering process. Grey and cursive letters indicate the star systems with contradicting stellar mass data.

### 3.1.2 Planet types, planetary configuration, and demography

The commonly used M vs. R plots were used to determine the type of planets associated with the candidate gas giants. Unlike the Solar system, none of the candidate star systems hosts terrestrial or super-Earth-type planets (Fig 1a). The planetary configurations contain gas giants and transitional planets/sub-Neptune types instead, located in the orbit of the terrestrials, referring to the Solar system as an analogy (Fig. 1b). Some of the associating planets orbit even inside the habitable zone of the star system, similar to the terrestrials in the Solar system (Table 1).

The plot in Figure 2 showing the possible relation between the stellar mass and the summarized mass of the known planets in the system has some characteristic components, outlined below:

i) Two groups of star systems can be observed in the plot, both with similar host star mass (it is given, considering the same stellar type used as a filtering parameter during selecting the candidate systems), but very different accumulated planetary mass falling between 1.2-2×10-2 M☉ and 1-8×10-2 M☉.

ii) For star systems with lower accumulated planetary mass, a significant relationship can be recognized between the stellar and accumulated planetary mass, assuming the connection between increasing stellar mass and the mass of planet-forming materials.

iii) Among the systems, the Solar system has the lowest accumulated mass in planetary bodies, even considering the mass of satellites, components of the Main Asteroid Belts, and Trans Neptunian Objects (Section 2.3). Along the Solar system, similar low accumulated planetary mass found in one of the oldest (8.5 Ga), six-planet containing HD 3445 system (Fig. 2; Table 1)

iv) The six-planet systems belong to the low accumulated planetary mass group.

### 3.2 Opening further discussions

The search for exoplanets with Jupiter and Saturn-like characteristics, targeting star and planetary systems that may host icy exomoons similar to the potential ocean- and life-harboring ones in the Solar system, showed surprising results. Such results can be summarized by the following thoughts, which hopefully trigger further discussions:

#### 3.2.1 About the number of potential candidates

There are 1374 gas giant-type exoplanets in the studied database, following the criteria of a gas giant, suggested by Zeng et al. [30] and Helled [31]. Among those exoplanets, thirteen planets, c.a. 1% of the gas giants group fulfill the criteria suggested in Section 2 (Methods) and make them Jupiter or Saturn-like analogs based on their physical parameters, the characteristics of their orbital parameters and their host star. Two of the 13 candidates must be excluded due to their controversial stellar age data. The others can be considered candidates for hosting icy satellites like exomoons.

There may not be any objective answer that such a c.a. 1% is high or low as a rate representing planets/planetary systems with putative exomoons, similar to the potential life-harbouring ones around Jupiter and Saturn, and among those the ones which may have an active role in the fertilization of the Universe and the Solar system from the viewpoint of Panspermia. As a subjective answer, having a higher possibility than zero already feels high enough to keep the hope alive and keep searching for further candidates or keep some of the existing ones under the radar of future exomoon studies.

### 3.2.2 About the missing terrestrial and super-Earth-type planetary neighbors

One of the surprising results of the data mining is that no rocky (terrestrial and super-Earth type) planetary neighbors appeared in the configuration of the candidate star systems. To find possible explanations, the star systems were classified by the allocation of the planetary orbits in the systems, regarding whether the planet is located in the Habitable Zone (HZ), the ice line, and in/out of the zone, which is analog to the distance between the ice line and Saturn ("Saturn-line") in the Solar system (for further information, please see Section 2).

Regarding the latter one, all the planetary neighbors of the candidate planets were inside the "Saturn line." Even the orbits of the ones outside of HZ do not reach further distances than the relative distance of Saturn from the ice line, like Uranus and Neptune in the Solar system.

Considering the outer limit of the Habitable Zone and the allocation of the ice line, most of the (inner) neighbors of the candidate gas giants orbit inside the habitable zone and the ice line of their star system. In addition, most of the planets inside the HZ are located in similar orbits to the terrestrial planets in the Solar system and belong to the group of transitional planets and sub-Neptune types (Table 2; Fig 1b). Such observation seems distinct in the case of six-planet-star systems, where the candidate planet is the sole planet orbiting outside the ice line, along with a group of planetary neighbors inside the HZ (Table 2; Fig. 2b). In the case of the three-planet-system, a transitional planet was found inside the HZ and a smaller gas giant beyond the ice line, orbiting inside the orbit of the outermost, candidate planet. Although it is challenging to come to clear conclusions due to the low number of planets, such a pattern appears in two-planet systems as well, with the candidate gas giant outside the HZ and ice line and one transitional planet/sub-Neptune inside, around the orbit of the expected terrestrials (HD 219828 and HD 140901 systems). There are some exceptional cases among the candidates, namely HD 66428, where a massive gas giant (HD 66428 b with 10.8 MJ) orbits between the ice line and the candidate planet (the four to five times smaller HD 66428 c planet). Other cases are HD 927088, where the additional gas giant orbits inside the HZ, and HD 50499, where the other gas giant is located between the outer limit of the HZ but inside the ice line. The lack of terrestrial and super-Earth-type planets in the candidate solar system may be explained by considering the following theories.

The transparent grey color areas I, II, and III mark: I) higher probability of material ejection from the oldest systems (HD 191939, HD 4203, and HD 34445) considering short ice shell formations on the icy exomoons; II) higher probability of material ejection from HD 21710 and HD 219828 exosystems with short ice shell formations on the icy exomoons; and III) higher probability of material ejection from younger exosystems (HD 140901, HD 156279 and HD 66428) with short ice shell formation (≤100 Myr). The green dashed line indicates the start of potential bioaerosol ejection from the older HD 191939, HD 4203, and HD 34445 exosystems if longer (3.5 Gyr) ice-shell formation was considered (Chapter 3.2.4). The area with a diagonal dotted pattern indicates the exosystems with the potential for material transport too early toward the Solar system. The dashed dark grey line marks the Solar system's formation, and the solid red line indicates the age of the first known fossil on Earth. The various transparent colors marks: light green - exosystems with a relatively low possibility of ejection and transport; dark green - exosystems with a relatively high possibility of ejection/and transport; red - exosystems with no possible transport before the appearance of life on Earth, and blue - the Solar system

The most naive explanation would be that current surveys of exoplanets are biased towards high-mass planets. Due to their physical parameters, transitional planets and gas giants are easier to find than terrestrial and even super-Earth-type planets. This may explain the missing additional planets in two-planet star systems (and certainly applies to putative icy exomoons [52-54].

In the studied dataset, 10% and 15% of the exoplanets belong to terrestrial and super-Earth types, respectively, preferably orbiting around low-mass stars ($M_{hostAVG}$: 0.82 and 0.88 $M_{\odot}$, respectively). The filtered database with candidate exosystems may represent the same ratio, where out of 12 systems (Solar system + the candidate exosystems), only one consists of rocky (terrestrial) planets (8.3%). In addition, there are six-planet star systems among the putative systems, consisting of planets located in the orbit of terrestrials in the Solar system, suggesting that no terrestrials (neither super-Earth type planets) have evolved (or evolved but “disappeared”?) during the evolution of the protoplanetary disk for various reasons, shortly summarized below. Putting aside the technical limitations, there are alternative explanations, possibly describing some of the factors causing the lack of rocky inner planets and the configuration of the exosystem. Regarding the initial phase of planet formation, the protoplanetary disk structure seems to be a significant factor in the demography of forming planets in the star system [55]. The evolution of structured (gapped) protoplanetary disks most likely results in the formation of gas giants in the exosystem. In contrast, the structure of disks with no gaps is more compact, which may favor the formation of rocky terrestrial and super-Earth-like planets.

Following the initial setting in material distribution in protoplanetary disks with various structures, additional processes influence the demography of the planets in the system.

The lack of protoplanetary material in the inner region of an exosystem can/could be one of the crucial segments of rocky planet formation. As has been discussed in various studies, protoplanetary disks may function as “conveyor belts,” transporting material from the outer regions toward the host star(s) [56]. Those inward-drifting pebbles (mm size, abundant dust aggregates) feed the growth of terrestrial and super-Earth-type inner planets [57], even from outside reservoirs from their protostellar cloud cores (e.g., tail-end accretion) [58]. In the case of the Solar system, such an inward shift of pebbles and their accretion resulted in the formation of planetesimals and later terrestrial protoplanets around Mars' orbit. Later, protoplanets migrated further inwards while growing and occupied their orbits, which now appear [57].

Based on simulations, the formation of outer planets with significant mass (e.g., gas giants more massive than Jupiter, may function as barriers and are capable of blocking the conveyor belt [59], and may result in pebble isolation at various locations (regarding the ice line) and the appearance of sub-Neptunes in the inner regions of the exosystem [60] (Fig. 3). Gas giants may block inward migrating super-Earth type planets (potentially it may apply to terrestrial size planets as well), turning them into future additional gas giant cores [61]. Such theories explain the appearance of transitional planets/sub-Neptunes in the inner exosystem of, e.g., HD191939 and HD 34445, and exosystems consist of two gas giants with significant mass differences between the two members (e.g., HD 66428). Regarding the former examples and the scenarios introduced in Bitsch et al. [60], the inner planets may be defined as dry and wet sub-Neptunes. In the case of the latter example, the appearance of gas giants with significant mass differences may indicate the accretion of inward migrating terrestrial and super-Earth-like planets in the “overgrown” member`s core.

### 3.2.3 About the two characteristic groups of planet associations

Some reasons for the lack of terrestrials and super-Earth-type planets may explain the appearance of the low and high accumulated planetary mass groups in the plot, introduced in Figure 2. The planets that belong to the high accumulated planetary mass group are members of two to three planet-containing systems, mainly consisting of gas giants with mass one magnitude bigger than Jupiter`s mass, such as HD 219828 c: 17 MJup,

HD 156279 b and c: 9.4 and 10 $M_{Jup}$ respectively, HD 66428 b: 10.8 $M_{Jup}$, and HD 204313 e: 15.2 $M_{Jup}$. The appearance of gas giants of such size may support some of the theories suggested above, namely, the gas giants with multiple Jupiter masses may block the inward migration of matter in general [59]. They can even turn inward, migrating super-Earth-type protoplanets into the core of other gas giants [61].

In contrast, the low accumulated planetary mass group is undoubtedly versatile. Various types of planetary configurations appear, such as the two six-planet containing exosystem, HD 191939 and HD 34445, with transitional/sub-Neptune inner planets and gas giants in the outer exosystem, the Solar system with unique demography including terrestrial, gas, and ice giant planets, and a group of two-planet systems, consisting mainly a duo of gas giants. The average mass of gas giants in this group is 2.8 $M_{Jup}$ with a maximum of 5.3 $M_{Jup}$ in the case of HD 140901 c. Regarding the theory of Sánchez et al. [59] about certain masses of gas giants blocking the dust/matter conveyor belt and influencing planetary formation in the exosystem, the suggested $M_{planet}$ = ~5 $M_{Jup}$ may represent some threshold, considering the mass of HD 140901 (5.3 $M_{Jup}$) and HD 204313 b (4.9 $M_{Jup}$). The former belongs to the low accumulated mass group, and the latter is a high accumulated mass group member. This may suggest two possible evolutionary paths for exosystems consisting of gas giant(s) five times bigger than Jupiter.

### 3.2.4 About the Panspermia approach

The proposed theory in the Introduction section assumed that theoretical icy exomoons of Jupiter and Saturn-like gas giants in exosystems may function as the incubators of life and fertilize certain parts of the Universe, including the Solar system, by microbial transfer via, e.g., asteroids. Considering the 3.465 Ga age of the possible earliest microfossils found on Earth so far [62]. Exosystems with age

$$Age_{sys} \geq T_{life\oplus} + T_{trav} \text{ (Ga)} \qquad \text{(Eq.3)}$$

can be considered as potential sources of life, where Agesys is the age of the exosystem, $T_{life\oplus}$ is 3.465 Ga, the oldest known lifeform on Earth [62], and $T_{trav}$ is the estimated traveling time to the Solar system.
Using the speed of the two known interstellar visitors in the Solar system, 1I/2017 U1 (Oumuamua; 87.3 km/s) and C/2019 Q4 (2I/Borisov; 48.61 km/s), the traveling time of potential complex organic molecules and microbial life hijacked asteroids from the candidate exoplanets were estimated. The travel time spams from the longest 1.64 Myrs (from HD 4203; 2I/Borisov speed) to the shortest 0.17 Myrs (from HD 140901; Oumuamua speed). Only some short speculation can be made regarding the survival of microbial life during the hundred-thousand to million-years-long transfer phase [1]. Recent experiments on ISS (International Space Station; Tanpopo mission) showed that microorganisms can survive a minimum of three years when exposed to outer space during interplanetary travel by using the surface cells killed by radiation as a protective layer (massapanspermia) [63]. Such a shield might allow safe travel toward the Solar system. However, our knowledge of possible survival strategies of a microbial community on an asteroid for such a long time is still minimal.

In addition to the appearance of life on Earth as the putative end point of the transfer phase referring to Panspermia theory, certain aspects regarding the evolution of icy satellites should also be considered while calculating the theoretical age of the source exosystem. There may be considerable time from the accretion of the exomoon-making materials and planetary differentiation to a geological period when simple lifeforms are expected. Although life can quickly spring measured on a geological timescale, certain environmental conditions must be fulfilled to allow it to happen. Such environmental conditions may need considerable time to appear following the accretion phase of exomoon evolution. Differentiation of an exomoon interior and the appearance of liquid water are possibly triggered by various heat sources, such as deposition of impact energy during accretion, radioactive decay, energy release through water-ice differentiation, and heat of mineral hydration reactions [64] and kept liquid even after the formation of the outermost ice-shell by mainly tidal heating and the heat from radioactive decay [65,66]. Based on the study of Schubert et al. [67], using Titan to explain the differentiation of icy satellites, it was estimated that the differentiation phase finished approximately 0.1 to 0.5 Ga following the accretion. The differentiation phase of a satellite (Europa, Jupiter) may connect to the so-called “hot phase” (c.a. 1.1 to 2 Ga), followed by the “cooling phase” (2 to 3.2 Ga) and the “equilibrium phase” (3.2 to 6.4 Ga), marked by fluctuating tidal dissipation rate and temperature [68]. Referring to the model of Schubert et al. [67], the surface ocean of Titan started to crystallize approximately around 3.5 Ga after the accretion. Such a process seems to be much faster in the case of Europa, where the crystallization of the ice shell is estimated to happen in 2 to 100 Myr [69]. The development of the ice shell feels necessary from the point of biological evolution, considering the strong magnetic field of the host planets of icy satellites of the solar system. Some of the icy satellites are exposed to radiation originating from the magnetic field of their host planet [70-73]. In the case of a lack of atmosphere or having only a thin envelope around the putative icy exomoons, the existence of the ice crust feels crucial to protect the evolving life in the subsurface ocean.

Referring to the results of the simulations [67,69], there is a 2 Myr to 3.5 Gyr time frame for the crystallization of such protecting ice shells, which needs to be considered when selecting the possible source exosystems in the case of a putative microbial transfer. Completing the condition above, the age of the candidate exosystem with gas giants hosting icy life harboring exomoons has to be:

$Age_{sys}-T_{sf} \geq T_{life\oplus} + T_{trav}$ (Ga) (Eq.4)

where $Age_{sys}$ is the age of the exosystem, $T_{sf}$ is the estimated length of shell formation (slow: 3.5 Gyr, and fast: 100 Myr and 2 Myr estimations; [67,69]), $T_{life\oplus}$ is 3.465 Ga, the oldest known lifeform on Earth [62], and $T_{trav}$ is the estimated traveling time to the Solar system.

Additional components of Panspermia transport should be considered. Bioaerosols were most likely ejected by asteroid impacts [1]. The probability of such collision is higher at the early period of the star system, considering a Late Heavy Bombardment-like period, in which regardless of its quasi-continuous, multiple cataclysms or single cataclysm nature, the planets experienced frequent impact events [74-76]. Calculating with the timing of the Late Heavy Bombardment period (c.a. >3.8 Ga) as analog, the highest probability of the ejection of bioaerosol is in c.a. 1 Gyr after the formation of the given star system. As a consequence of the long, 3.5 Gyr formation period of some icy satellites [67], the evolving life below their ice shell may/might miss the period with a higher chance of ejection and transportation.

Considering the transportation length, there is a down part of the ejection of bioaerosols during a Late Heavy Bombardment-like period. Asteroids hurtling toward the Solar system following the asteroid bombardment period in the old candidate exosystems (8-9 Ga; Table 1, and Fig. 3) may arrive too early to the assumed location of the Solar system, even calculating with the longest, 1.64 Myr, traveling time.
Such a paradox means that there is a higher probability of early bioaerosol ejection due to the putative Late Heavy Bombardment-like period in the candidate star system. Still, the transported material may arrive too early, along with the scenario having long satellite formation, which postpones the formation of life in a relatively permanent planetary system and significantly decreases the chance of bioaerosol ejections due to the low number of asteroid impacts, leaving a relatively narrow window for fertilization of the Solar system.

Considering long ice shell formation around the source icy exomoons, only three candidate exosystems are left from the list with older age: HD 191939, HD 4203 and HD 34445. The bioaerosols of the icy satellites of the candidate planets in those exosystems might have a lower chance of asteroid impacts and ejection. In addition, in the case of the older candidate systems, the bioaerosol might arrived too early and the Solar system had not been formed yet to be fertilized (Fig. 3). It might happened especially if shorter ice shell formation time was applied, resulting higher probability of bioaerosol ejection due to the higher asteroid impact rate during the putative Late Hevy Bombardment-like period, but lower chance for arrival in time to the Solar system. Younger systems, such as HD 217107, HD 219828, HD 140901, and HD 156279, maybe better candidates if a short, 100 Myr long ice shell formation period is applied along with the higher probability of bioaerosol ejection and transportation. Ejectiles from HD 217107 and HD 219828 still have some chance of arriving too early at the Solar system's location. Considering the minimum 2 Myr ice shell formation estimation [69], HD 66428 can be added to the list of potential extraterrestrial life source systems at a “reachable” distance from the Solar system.

## 4. CONCLUSION

The search results for exosystems with Sun-like host stars and having Jupiter and Saturn-like planets and Solar system-like planetary demography showed somewhat surprising results. In the over 5000 exoplanets containing list, c.a. a dozen fulfilled the filtering criterion considering the Solar system's main characteristics and the dataset's limitations (i.e., only basic parameters could be used to maximize the number of candidate planets with necessary data availability).

The solar system feels unique based on the planetary demography of the candidate exosystems. No other systems with rocky planets were found. This result may support the introduced theory, i.e., that the icy satellites of commonly appearing gas giants more likely functioned as source planets in the fertilization of the Universe (Panspermia) than rocky planets.

Another factor to consider during the reconstruction of the transportation of microorganisms is its timing. Ejection of microorganisms is more likely during the earlier evolutionary stage of the star system, when the frequency of asteroid impacts, which may be capable of ejecting the material out of the gravity well, was much higher than at a later time. Unfortunately, considering more extended differentiation estimations for those satellites, the source body might not have any microorganisms or a fully developed planetary interior and surface. There is another difficulty in counting. Even if there were some microorganisms to transfer, a too-early ejection toward the Solar system would not been able to fertilize a system that had not existed yet.

The search for systems with Jupiter and Saturn-like planets and comparing the candidates with the Solar system were somewhat unsatisfying from the angle of the theory about the role of icy exomoons in the fertilization of the Universe. Although it seems that planetary demography with dominantly gas giants was/is more common, i.e., their icy satellites could function as a source (if they harbor life), the critical factor in a direct exosystem to Solar system Panspermia transport turned out to be the timing.

## STATEMENTS AND DECLARATIONS

Acknowledgments. We want to thank the anonymous reviewers of this article for taking the time and effort to review the manuscript. We appreciate their valuable comments and suggestions, which helped improve its quality. This research has made use of the NASA Exoplanet Archive, which is operated by the California Institute of Technology, under contract with the National Aeronautics and Space Administration under the Exoplanet Exploration Program.

Funding. No funding were used during the execution of the research.

Conflict of interest. The author has no conflicts of interest to declare.

Supplementary Information. Data used in the study will be provided from the author upon request.

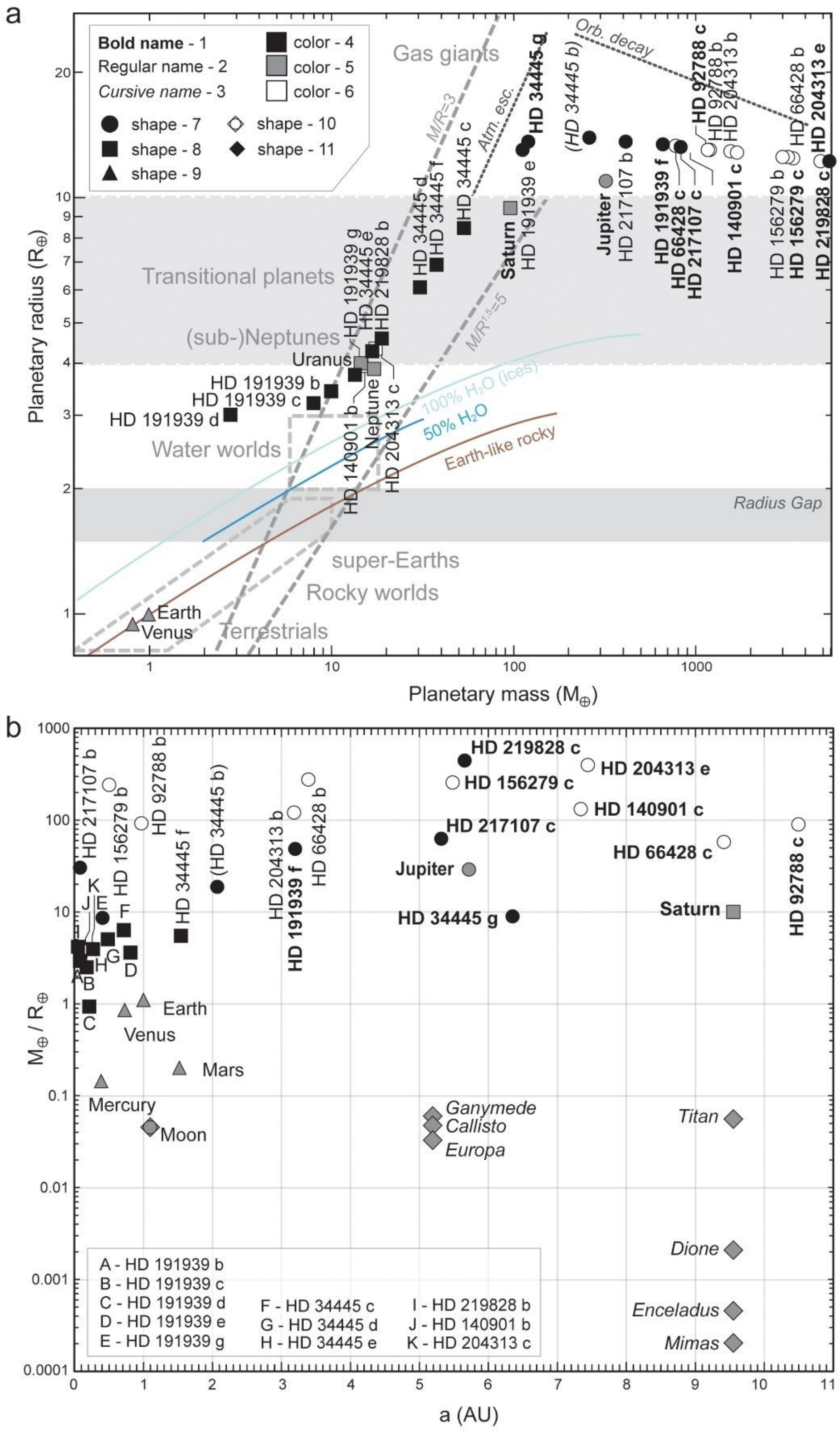

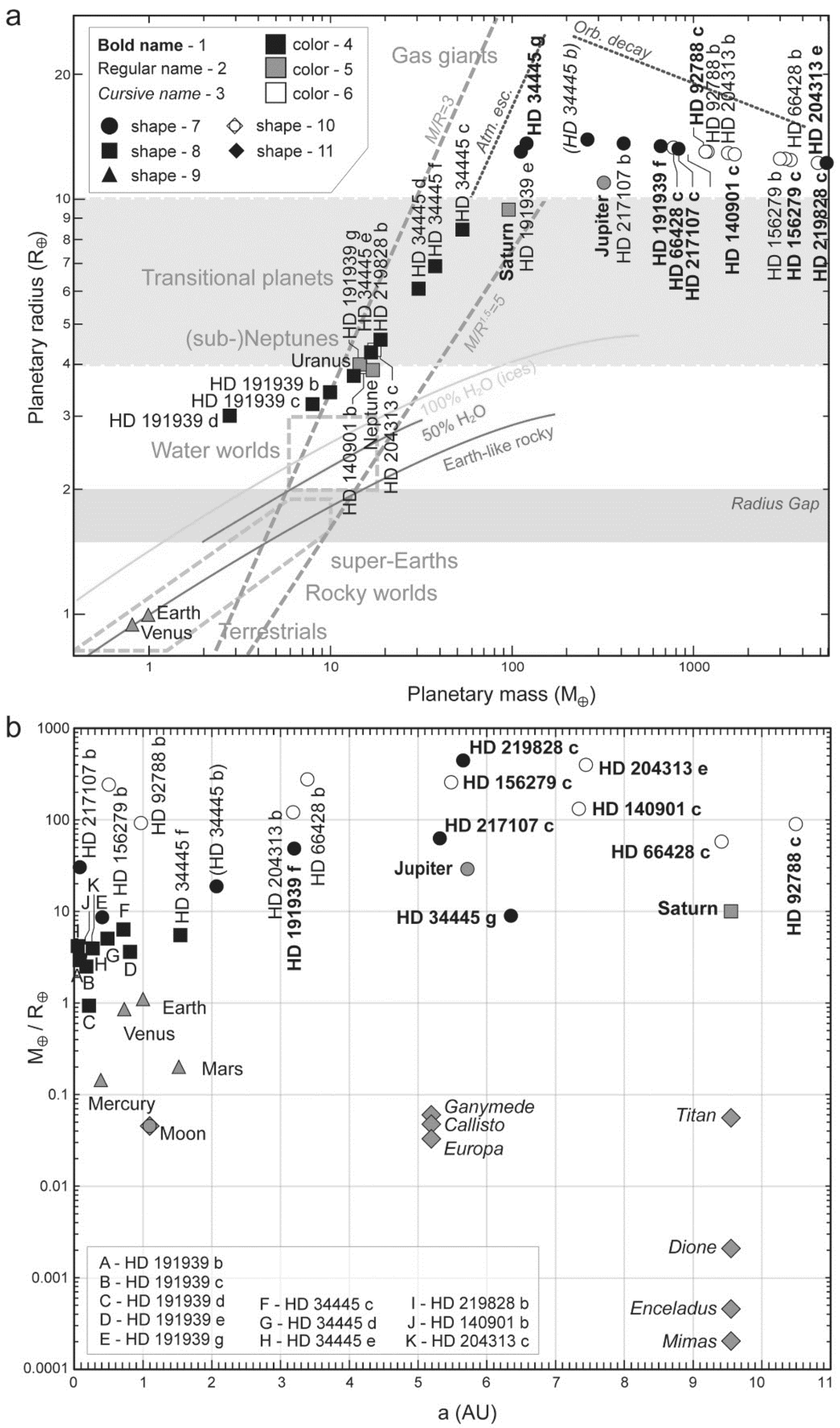

**Figure 1.** The classification of the planets of the candidate icy exomoon-bearing systems by MR plot (a) and the planetary demography of the systems and its comparison to the Solar system. 1 – candidate Jupiter and Saturn analog gas giants; 2 – additional planets in the star system; 3 – some of the known possibly ocean-bearing icy satellites in the Solar system; 4 – exosystems older than the solar system; 5 – Solar system planets; 6 –

exosystems younger than the Solar systems; 7 – gas giants; 8 – transitional planets/sub-Neptunes; 9 – rocky planets (terrestrials and super-Earth types); 10 – Moon-like satellites; and 11 – icy satellites. The base plot of Figure 1a is adapted from Zeng et al. [30].

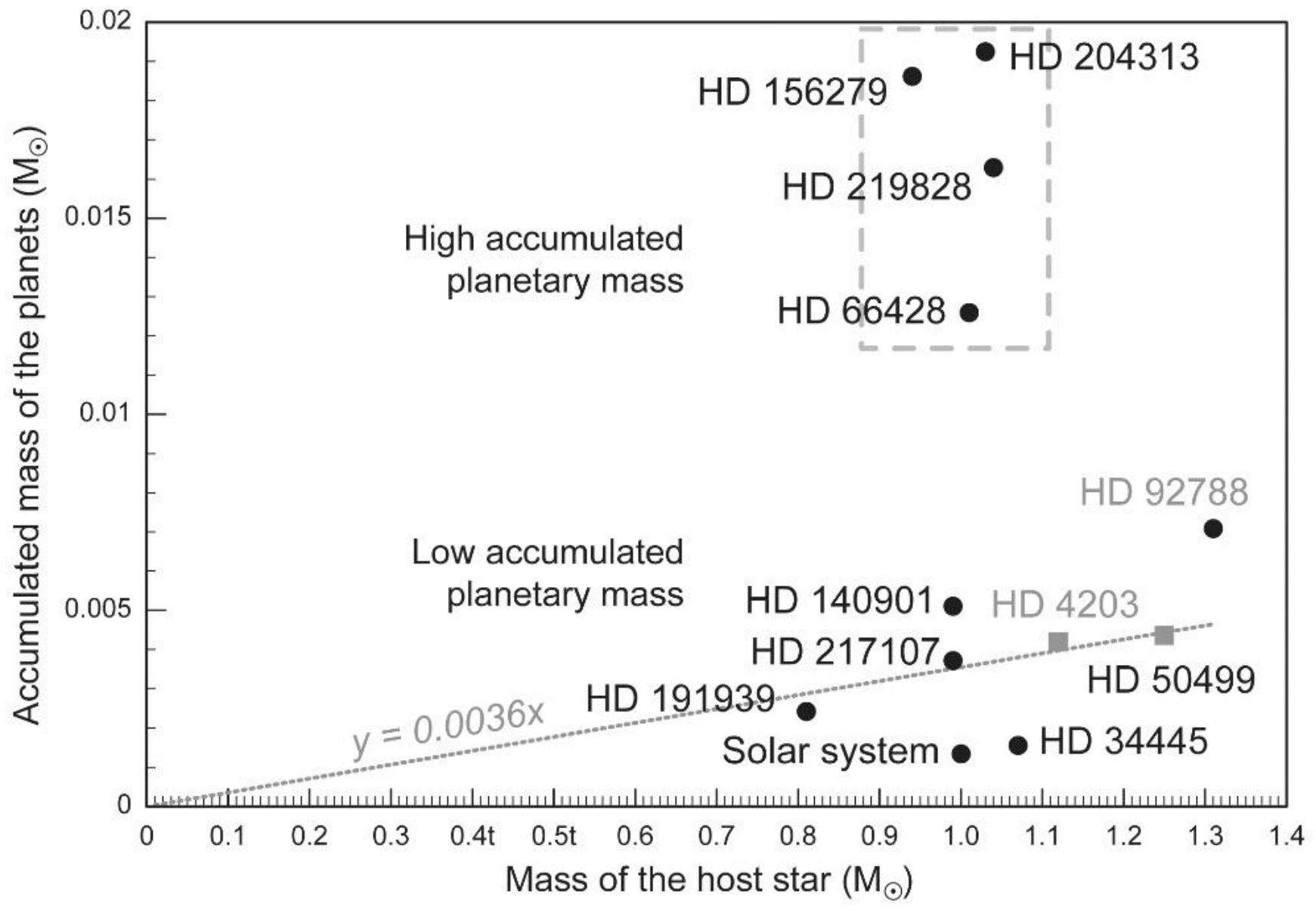

**Figure 2**. Revealing the statistical relationship between the host stars and the planet's mass in the candidate exosystems. The Pearson correlation coefficient and the results of the significance test (t-test) for the trend recognized between the mass of the host star and the accumulated planetary mass is the following: $r = +0.9$, $p(a) < 0.05$ (level of significance = 0.71), $n = 8$ (number of records). The grey squares indicate the systems with contradicting stellar mass (see Table 2, and section 3.1.1) where the average was calculated. The dashed line marks the systems with higher accumulated planetary mass.

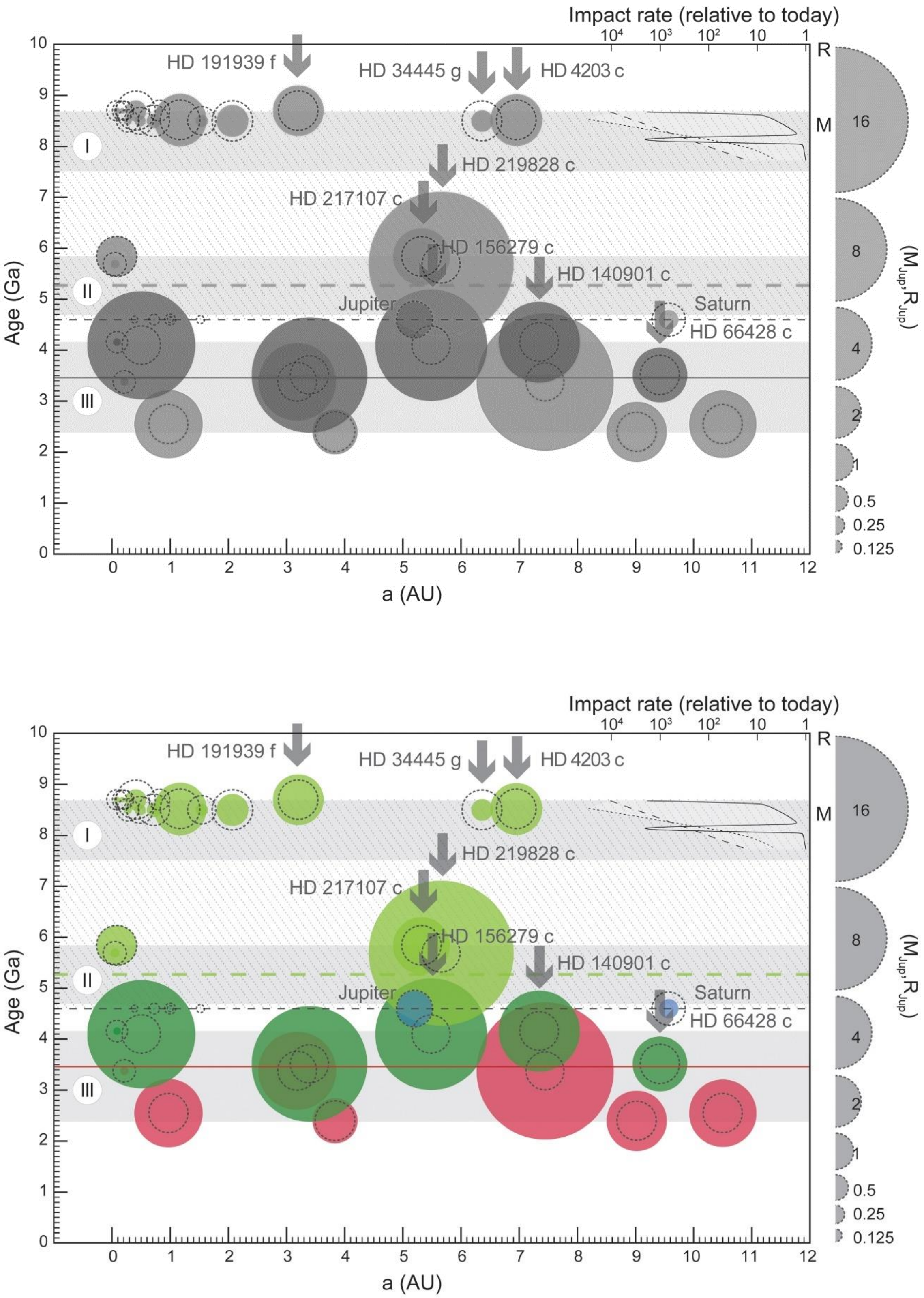

**Figure 3**. The age and configuration of the known planets in putative, icy exomoon-bearing Jupiter-like gas giant consisting of candidate exosystems. The size of the bubbles indicates the mass of the planets in $M_{Jup}$. The dotted, dashed, and solid black curves in the upper right corner of the plot show the estimated impact ratio in the case of Late Heavy Bombardment and are used as an indicator of higher probability bioaerosol ejection in space

(steady decrease impact flux: dotted and dashed lines, and single cataclysm - sensu stricto Late Heavy Bombardment: solid line scenarios) [51]. I.e. the curves may show the changing tendency in the probability of bioaerosol ejection and transportation in the case of various exosystems in the period marked by the transparent grey horizontal regions.